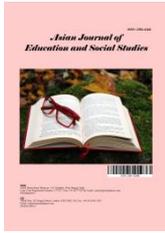

# Econometrics Modelling Approach to Examine the Effect of STEM Policy Changes on Asian Student's Enrollment Decision in USA

**Prathamesh Muzumdar [a\*], George Kurian [b\*], Ganga Prasad Basyal [c] and Apoorva Muley [d]**

[a] *The University of Texas at Arlington, USA.*
[b] *Eastern New Mexico University, USA.*
[c] *Black Hills State University, USA.*
[d] *People's University, Bhopal, India.*

***Authors' contributions***

*This work was carried out in collaboration among all authors. All authors read and approved the final manuscript.*

*Original Research Article*

## ABSTRACT

Academic research has shown significant interest in international student mobility, with previous literature primarily focusing on the migration industry from a political and public policy perspective. For many countries, international student mobility plays a crucial role in bolstering their economies through financial gains and attracting skilled immigrants. While previous studies have explored the determinants of mobility and country economic policies, only a few have examined the impact of policy changes on mobility trends. In this study, the researchers investigate the influence of immigration policy changes, particularly the optional practical training (OPT) extension on STEM

___

*Corresponding author: Email: prathameshmuzumdar85@gmail.com, george.kurian@enmu.edu;*

programs, on Asian students' preference for enrolling in STEM majors at universities. The study utilizes observational data and employs a quasi-experimental design, analysing the information using the difference-in-difference technique. The findings of the research indicate that the implementation of the STEM extension policy in 2008 has a significant effect on Asian students' decisions to enroll in a STEM major. Additionally, the study highlights the noteworthy role of individual factors such as the specific STEM major, terminal degree pursued, and gender in influencing Asian students' enrollment decisions.

## 1. INTRODUCTION

Over the past twenty years, there has been a notable increase in student mobility towards higher education institutions in developed countries, as highlighted in the work of De Wit et al. (2013). This trend facilitates the exchange of knowledge, capital, and goods among diverse groups of people, particularly in comparison to less-educated immigrants [1]. Previous studies have examined the positive financial effects that high-skilled immigration brings to the economy [2]. The demand for high-skilled labor and skill-intensive job roles has mutually driven organizations to attract skilled workers and invest in training their existing workforce to upgrade their skills [1]. This proactive approach enables organizations to adapt to the evolving technological needs [3]. The influx of international students in the USA has significantly influenced the labor market for high-skill workers [1,4], leading to substantial impacts on this specific segment of the labor force [5].

The mobility of international students has sparked immigration policy debates in the USA, and over the last two decades, changes in immigration policies have influenced the job role preferences of international students [6]. Starting from the early '90s, the USA has experienced a significant influx of skilled workers from foreign countries, particularly with many Asian workers taking up high-end tech jobs [5]. This influx of foreign workers was accompanied by a rise in the number of international students arriving in the USA, seeking higher education and job opportunities [1]. Notably, these international students have followed the same trend of opting for tech job roles in the industry, likely influenced by the work patterns established by their predecessors [7]. Given the consistency of international students' preference for specific professions, this situation presents an ideal opportunity to examine the impact of immigration policies on their job role preferences.

The focus of this study is to analyze the impact of the STEM extension policy change on Asian students' decisions to enroll in a STEM major program, with reference to the research conducted by Hawthorne in 2010. The study aims to compare two distinct groups: Asian students holding a student visa and US-born or permanent resident US students. By examining these two groups, the research aims to investigate whether the enrollment decisions of Asian students on student visas were influenced by the changes in the STEM extension policy implemented in 2008. The comparison will be made between Asian students with student visas and those who have US citizenship or permanent residency.

## 2. BACKGROUND AND HYPOTHESIS

### 2.1 Background

Since the 1970s, higher education in the USA has increasingly attracted foreign students, leading to a significant rise in student numbers until the beginning of the twenty-first century [1]. The mobility of international students has not only impacted the structure of the USA's higher education system but has also introduced a considerable pool of skilled labor, particularly benefiting the IT labor market [8].

The immigration policies of the USA have been influenced by the trends in foreign student mobility, resulting in a process where employers in the high skilled job market prefer sponsoring work permit-related visas [9,10]. In this context, the changes in STEM-related immigration policies have been implemented to support Asian student visa holders in extending their stay in the USA by an additional 17 months after completing their 12 months of Optional Practical Training (OPT). This extension enables these students to explore job opportunities in the US job market.

Considering the above background, the study proposes three research questions.

The STEM policy change implemented in 2008 aimed to specifically attract students to enroll in STEM programs, with the goal of boosting higher education revenue and cultivating skills for the technology job market in the USA [11]. These policies were designed to entice foreign students to join programs with a higher likelihood of leading to success in tech-related professions [12,13]. STEM programs predominantly focus on technical fields, requiring students to concentrate on courses related to engineering and science [14,15]. It has been observed that many Asian countries exhibit a strong inclination towards engineering and science as preferred pathways for securing jobs in both local and international markets [16]. Previous literature has extensively examined immigration policy concerning students in general and those who opt for part-time enrollment, both of whom fall into the category of immigrants [17,18]. These studies have shed light on various aspects of immigration policies pertaining to student populations.

In previous research, immigration policies have been extensively studied from an economic and job market perspective [19]. Additionally, there have been studies exploring immigration patterns concerning segmented minorities and their concentration in specific skill markets [20,21]. However, to date, there has been a gap in research when it comes to focusing on students from Asian ethnicities and their inclinations towards particular fields of science in higher education. This study aims to address this gap by not only investigating Asian students' preferences for STEM programs but also their aspirations to remain in the USA by entering the job market [22,23]. The study recognizes that the STEM policy allows for an extended 17-month period for Asian students to search for job opportunities, and this aspect will be taken into account in the research [24]. With this context in mind, the study formulates three research questions to delve into the specific dynamics of Asian students' choices and aspirations within the realm of higher education and the job market.

Research question 1 aims to investigate the impact of STEM extension immigration policies on the enrollment decisions of Asian students [25,26]. This question is at the heart of the study, as it seeks to understand the specific influence of the STEM extension policy on the preferences of Asian students holding student visas [27]. By focusing on the policy as the primary determinant, the research will analyze how the availability of an extended 17-month period for Optional Practical Training (OPT) affects Asian students' choices when it comes to enrolling in STEM majors at universities. The question delves into the policy's role in shaping the decision-making process of Asian students and how it may influence their inclination towards pursuing STEM programs.

**RQ1:** Does OPT extension policy affect the immigrant Asian students' decision to enroll in a STEM program when compared to American-born Asian students?

**Table 1. Literature review**

| Authors | Topic | Findings | Type |
| --- | --- | --- | --- |
| Dorantes et. al (2018) | Immigration policy | Effects of OPT policy changes on students | Quantitative |
| Ruiz and Budiman (2018) | Immigration policy | STEM major statistics | Descriptive |
| Demirci (2019) | Immigration policy | Large policy impact | Quantitative |
| Klimaviciute (2017) | Labor policy | Effects of STEM policy on labor markets | Quantitative |
| Rosenzweig (2006) | Study abroad and immigration | Decision modelling of student's enrollment decision | Quantitative |
| Bound et al. (2015) | US IT labor market | US degree important pathway to US IT labor market | Quantitative |
| Clemens (2013) | Wages and labor market | Premium wages influencing IT labor markets | Quantitative |
| Bound et al (2015) | Immigration labor market | International student market share in labor market | Quantitative |
| Kato and Sparber (2013) | Student decision making | Attributes influencing decision making | Quantitative |

Research question 2 aims to explore the effects of other factors that influence the enrollment decisions of Asian student visa holders. While immigration policy plays a crucial role, it is acknowledged that enrollment decisions are influenced by multiple factors, and immigration policy alone cannot solely determine the course of enrollment [28]. The study will take into account various extraneous variables and examine their significance in shaping Asian students' decisions when choosing their majors at universities. These factors could include but are not limited to personal interests, career aspirations, financial considerations, cultural influences, and educational background [29,30]. By considering these additional determinants, the research aims to provide a more comprehensive understanding of the complex interplay of factors affecting Asian students' enrollment decisions in STEM programs. It will highlight the multi-faceted nature of the decision-making process and how various aspects come together to influence their choices beyond just immigration policies.

> **RQ2.** What other factors significantly influence immigrant Asian students' decision to enroll in a STEM program?

Research question 3 focuses on investigating the heterogeneous effects of a specific STEM major on the decision-making process of Asian student visa holders when it comes to enrolling in STEM programs. Unconditional heterogeneity refers to the presence of diverse characteristics among individuals that are not predictable over time. In the context of this study, the unpredictable nature of STEM majors comes into play because they qualify each Asian student visa holder for an extended stay in the USA under the STEM extension policy. The research aims to understand how the choice of a particular STEM major influences Asian students' decisions to enroll in STEM programs, considering the various individual characteristics and preferences that may come into play. By studying the effect of different STEM majors on enrollment decisions, the research seeks to identify potential patterns and variations in Asian students' choices [31,32]. This analysis will contribute to a more nuanced understanding of the impact of specific STEM fields on their educational preferences and plans for their stay in the USA.

> **RQ3.** Does a particular STEM major influence this decision more than other majors?

The three research questions focus on understanding the influence an immigration policy has on determining an enrollment decision taking into consideration the student's willingness to continue in the USA job market.

## 2.2 Theory Development and Hypothesis

### 2.2.1 Hypothesis 1

The first hypothesis of the study proposes that the STEM extension policies significantly influence the enrollment decisions of prospective Asian students on student visas [33]. The hypothesis suggests that these policies, which allow students to extend their stay in the US for practical work experience and skill development, play a crucial role in shaping the preferences of Asian students when choosing their majors at higher education institutes in the US. The hypothesis also suggests that Asian students may be more inclined to enroll in STEM majors, as these programs qualify them for the STEM extension, providing them with the opportunity to gain practical work experience and potentially secure better career prospects in the US job market [34]. By testing this hypothesis, the study aims to gain insights into the impact of STEM extension policies on Asian students' enrollment decisions and whether these policies act as a significant driver for their focus on STEM majors during their studies in the USA.

> **H1.** OPT extension provision positively affects the immigrant Asian students' decision to enroll in the STEM majors compared to US citizen Asian students

### 2.2.2 Hypothesis 2

Asian students holding student visas contribute significantly to the substantial revenues generated in the US higher education market. Research by Muzumdar et.al (2020) suggests that most of these enrolments follow a discernible migration pattern. This pattern encompasses students originating from specific income groups, family backgrounds, educational histories, and more [35]. Given that a majority of Asian student's hail from developing countries, it becomes essential to explore the role of gender in their enrollment decisions. The study aims to ascertain whether students from certain genders exhibit a preference for enrolling in STEM programs. By examining the relationship between gender and the choice of STEM majors among Asian students on student visas, the research seeks to uncover potential gender-related patterns in enrollment decisions. Understanding this aspect will provide valuable insights into how gender dynamics influence Asian students' academic choices, particularly within the realm of STEM fields in the USA.

**H2**. Being a male immigrant Asian student positively influence a students' decision to enroll in the STEM majors

### 2.2.3 Hypothesis 3

Terminal degrees have been regarded as a means to enhance one's skillsets and knowledge in a specific major of focus [6,36]. Pursuing such degrees is also seen as an investment in personal development, driven by higher aspirations [20]. A Ph.D. represents one such terminal degree, demanding a commitment of five years to develop specialized skillsets. For STEM majors, obtaining a Ph.D. can facilitate the practical application of acquired skillsets post-graduation, whether in industry or academia [37]. The 17-month extension further enables Asian students on visas to augment their skillsets in their chosen fields, providing opportunities in both industry and academia following graduation [38]. Within this study, it becomes essential to explore the role of enrolling in a Ph.D. program after the policy's implementation, as it allows Asian students on visas to substantiate their five-year commitment to the program.

**H3**. Option to enroll in a Ph.D. degree positively influence a students' decision to enroll in the STEM majors

The three hypotheses explored in this study investigate the impact of a policy change on a specific ethnic group, which constitutes the largest proportion of students enrolling in higher education in the USA [39]. By examining the policy change, the research aims to gain insights into how new policies aimed at enhancing the job economy can influence the enrollment rates in specific STEM majors. The data and analysis, utilizing an econometric modelling technique, contribute to the development of the model and provide answers to the research questions [22]. Overall, this approach helps to comprehensively understand the relationship between policy changes, student demographics, and enrollment trends in STEM fields.

## 3. DATA

The data for this research was gathered from the National Survey of College Graduates (NSCG) spanning the years 2008 to 2016, thereby encompassing an equal balance of eight years before and after the policy implementation in 2008. The study's sample size consisted of 623,086 respondents. To ensure data integrity, the sample was sorted and cleaned for each unique respondent ID.

The variables considered in this research are as follows:

## 4. METHODOLOGY

### 4.1 Experimental Design

The experimental design employed in this research was a 2 x 2 design. The treatment group consisted of Asian students on student visas who were enrolled in STEM programs. On the other hand, the control group comprised Asian students who were either US citizens or permanent residents and were also enrolled in STEM programs. The two conditions examined were the STEM enrollment data for the pre and post STEM extension policy change periods [40]. The observational data collected was then organized and sorted for the values corresponding to the variables taken into consideration in the benchmark model (equation 1). This allowed for a comprehensive analysis of the impact of the STEM extension policy change on Asian students' enrollment decisions within the specified conditions.

### 4.2 Difference-In-Difference (DID)

In this research, the Difference-in-difference (DID) econometrics quantitative method is applied to a 2x2 experimental design using observational data. Since the data is archival and not behavioral, analyzing and interpreting it using a traditional experimental design is challenging. However, the DID technique serves as a valuable alternative, simulating an experimental setup for observational data by creating a dummy variable for the control group [15]. Specifically, the study employs a benchmark model to investigate the influence of the STEM extension policy on Asian students' decisions to enroll in STEM major programs while taking into account their immigration status [41]. The DID approach enables researchers to compare the changes in outcomes between the treatment group (Asian students on student visas in STEM programs) and the control group (Asian students who are US citizens or permanent residents in STEM programs) over time. By utilizing this methodology, the study can effectively assess the causal impact of the STEM extension policy on Asian students' enrollment decisions while controlling for the differences in their immigration statuses. This analysis aids in understanding the true effect of the policy change on the enrollment patterns of Asian students in STEM majors.

$$Y_{i,v,e,t} = \alpha + \beta_1 STEM_{v,e} + \beta_2 X_i + \delta_c + \delta_m + \delta_l + \varepsilon_{i,v,e,t}$$
equation (1)

In the benchmark model, the dependent variable, denoted as $Y_{i,v,e,t}$, represents an Asian student's enrollment in a STEM major program. For a particular Asian student "i" who enters with a student visa "v" and enrols in a specific calendar year "e," the variable takes a value of 1 if the student possesses a terminal degree in a STEM major at a given year "t." If the student does not have a terminal degree in a STEM major at that time, the variable takes a value of 0.

The coefficient of interest in this model is denoted as β1, which quantifies the impact of the STEM extension policy on the likelihood that Asian students on student visas would choose to enroll in STEM major programs [42]. By analysing the magnitude and significance of β1, the study can discern the extent to which the STEM extension policy influenced Asian student visa holders' decisions regarding their enrollment in STEM majors.

The vector Xi in the model incorporates various individual-level characteristics, such as age, gender, marital status, and terminal degree. Additionally, the model includes immigration status-related fixed effects (δc) to account for Asian students with different visa statuses (Student visa vs US citizen) and their preferences towards STEM majors. Another fixed effect (δm) is considered in the model to capture the preferences for specific STEM majors, which remain constant over time. This fixed effect is distinct from the dependent variable (DV) since STEM majors vary in number, and the enrollment in general does not define preferences towards a particular major.

Moreover, the model takes into consideration the fixed effects of enrollment year (δl). This accounts for the influence of the year of enrollment, as it plays a crucial role in determining job opportunities in the US labor market. By incorporating these fixed effects, the model aims to control for the impact of various factors on Asian students' enrollment decisions in STEM major programs, providing a comprehensive analysis of the policy's effect.

Probit regression is employed in this study to analyze the data, considering the binary nature of the dependent variable. The presence of conditional heteroskedasticity poses challenges as the level of volatility cannot be predicted accurately. Stem fields in the data contribute to conditional heteroskedasticity, given the difficulty in predicting immigrant Asian students' intentions to enroll in specific STEM majors.

The objective of this research is to predict the impact of the STEM extension policy not only on the overall enrollment into STEM programs but also on enrollment into particular STEM majors. To address the difficulties posed by heteroskedasticity, the study has chosen to use Probit regression [43]. This method accounts for the binary nature of the dependent variable and helps overcome the challenges associated with conditional heteroskedasticity, enabling a more robust analysis of the policy's influence on Asian students' enrollment decisions in both STEM programs and specific STEM majors.

**Table 2. Variable list**

| No | Variable | Definition | Authors |
|---|---|---|---|
| 1 | Program enrollment | Student enrolled in a program (general) | Dorantes et. al (2018) |
| 2 | Race | Race | Rosenzweig (2006), Dorantes et. al (2018) |
| 3 | Immigration status | Visa status, citizenship, permanent residency | Klimaviciute (2017), Dorantes et. al (2018) |
| 4 | Gender | Gender | Dorantes et. al (2018) |
| 5 | Year surveyed | Year in which the survey was carried out | Dorantes et. al (2018) |
| 6 | Terminal degree | Masters, doctoral, bachelors | Dorantes et. al (2018) |
| 7 | STEM major | Major | Ruiz and Budiman (2018) |

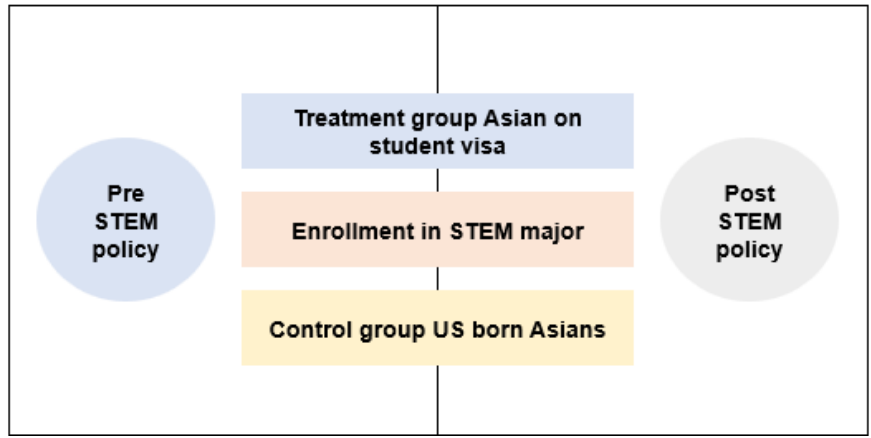

**Fig. 1. Research Design- Quasi experimental design**

## 5. DID ASSUMPTIONS

The difference-in-difference (DID) method serves as a valuable tool for analysing observational data by emulating experiments. Conducting field manipulations for observational data is challenging, as the data is pre-collected by a third-party entity and not directly controlled by the researchers [44]. To overcome this limitation, the DID method generates treatment effects by introducing dummy variables. In order to use the DID method for analysis, a key assumption must be verified: the treatment group and control group followed a parallel trend. This means that, in the absence of the treatment (i.e., policy change), both groups would have exhibited similar trends over time [45]. By comparing the changes in outcomes between the two groups before and after the treatment, the DID method enables researchers to estimate the causal impact of the treatment, making it a valuable approach for examining the influence of policy changes on observational data.

Indeed, if the trend of the share in STEM majors for Asian students (US citizens) and students on a student visa was parallel before the year 2008 but changed significantly after the policy change took place, it suggests that the assumption of parallel trends is supported.

When analysing the data using the difference-in-difference method, the presence of parallel trends is crucial [10]. It indicates that any differences observed in the trends after the policy change can be attributed to the policy itself, rather than pre-existing divergences in trends between the two groups. This strengthens the validity of using the difference-in-difference approach to estimate the causal impact of the policy change on Asian students' enrollment decisions in STEM majors and helps draw more reliable conclusions from the analysis.

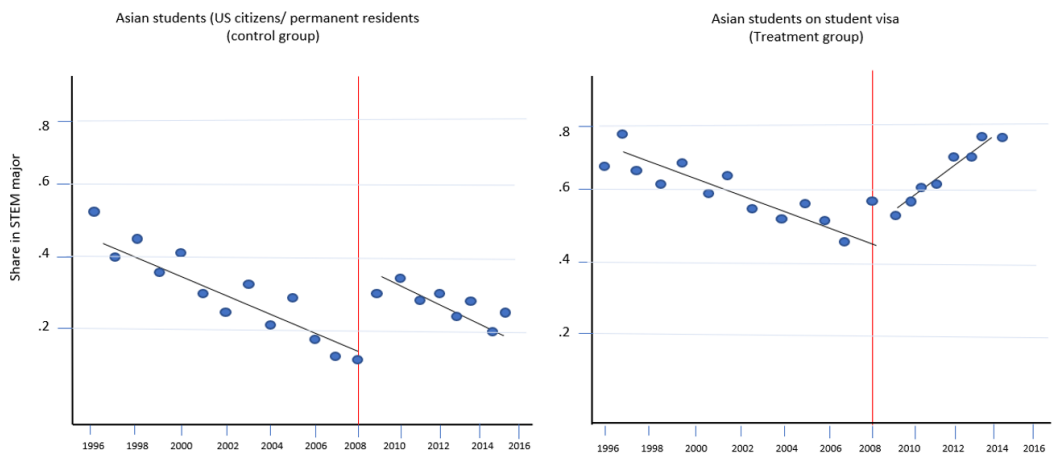

**Fig. 2. DID assumption**

## 6. RESULTS

In this study, the difference-in-difference (DID) econometrics quantitative method was utilized to examine the impact of the STEM extension policy, which was implemented by the USCIS in 2008. Probit regression was employed to analyze the causal relationship and quantitatively measure the significant impact of the independent variable of interest, namely, the STEM extension policy, on the dependent variable of STEM major enrollment.

Table 3 presents the results of the variables estimated in the analysis, as presented in the benchmark model (equation 1). This table likely includes the coefficients and statistical significance of various variables considered in the model, including the STEM extension policy, immigration status, enrollment year, and other individual-level characteristics.

The results in Table 3 provide valuable insights into the influence of the STEM extension policy on Asian students' decisions to enroll in STEM major programs, while controlling for other relevant factors. The statistical significance of the coefficients helps determine the magnitude and direction of the effects, thereby offering meaningful conclusions about the impact of the policy change on Asian students' STEM major enrollment decisions.

The results of the analysis strongly supported the main hypothesis. The variable of interest, STEM extension, was found to be statistically significant in a positive way. Specifically, the STEM extension policy significantly increased the likelihood of enrolling in a STEM degree by a substantial 48 percentage points. This finding indicates that the policy change had a significant impact on Asian students' decisions to pursue STEM majors, providing evidence of its effectiveness in encouraging STEM enrollment.

Furthermore, all the variables corresponding to the vector variable X were also found to be significant. According to the results in Table 3, males were found to be 16.8% more likely to enroll in a STEM major compared to females. Moreover, those planning to enroll in doctoral degrees were 18.6% more likely to choose a STEM major over those with a bachelor's degree, while those planning to enroll in master's degrees had a 42.5% higher likelihood of enrolling in a STEM major compared to those with a bachelor's degree.

Additionally, age appeared to be an important variable in the decision-making process, as it negatively affected the likelihood of STEM major enrollment. This suggests that younger students were more inclined towards enrolling in STEM programs compared to older students.

Overall, the findings strongly indicate that the STEM extension policy had a significant and positive influence on Asian students' enrollment decisions in STEM major programs, while other individual-level characteristics, such as gender, degree plans, and age, also played a significant role in shaping their choices.

**Table 3. Probit regression**

| No | Variable | Student visa Asian Coefficients |
|---|---|---|
| 1 | STEM extension | 0.48*** |
| 2 | Age | -0.028*** |
| 3 | Age squared | 0 |
| 4 | Male | 0.168*** |
| 5 | Married | 0.025*** |
| 6 | MS degree | 0.425*** |
| 7 | PhD degree | 0.186*** |
| 8 | Immigration status FE | Y |
| 9 | STEM Major FE | Y |
| 10 | Enrollment year FE | Y |
|  | Observations | 623,086 |
|  | R squared | 0.786 |

In nonexperimental studies, it is commonly assumed that error terms are uncorrelated with each other and with any independent variable. However, in reality, this assumption may not hold true, making it unrealistic in certain cases. To address potential confounding influences from a confounding variable, researchers often conduct robustness tests to ensure the reliability of their findings.

In this study, a robustness test was performed under four conditions, as indicated in Table 4. The first condition involved adding other races to the control group to assess their influence on the significance and coefficient estimate of the "STEM extension" variable. The results showed a slight increase in the coefficient and a higher R-squared value. However, this increase was primarily due to the significant increase in sample size resulting from the inclusion of other races, particularly Asian students from China and India, who heavily dominated the share of students enrolling in programs on a student visa.

**Table 4. Robustness check**

| No | Variable | Add other races in control group | Exclude China from sample | Exclude India from sample | Drop recession years |
|---|---|---|---|---|---|
| 1 | STEM extension | 0.52*** | 0.44*** | 0.46*** | 0.56*** |
| 2 | Age | -0.036*** | -0.026*** | -0.028*** | -0.032*** |
| 3 | Age squared | 0 | 0 | 0 | 0 |
| 4 | Male | 0.176*** | 0.162*** | 0.164*** | 0.172*** |
| 5 | Married | 0.032*** | 0.021*** | 0.022*** | 0.026*** |
| 6 | MS degree | 0.568*** | 0.422*** | 0.425*** | 0.464*** |
| 7 | PhD degree | 0.226*** | 0.182*** | 0.184*** | 0.188*** |
| 8 | Observations | 878,021 | 252,226 | 374,322 | 488,486 |
| 9 | R squared | 0.822 | 0.772 | 0.778 | 0.780 |

Conditions two and three involved excluding Chinese and Indian students, respectively, from both groups. The results showed a minor decline in the STEM extension coefficient and R-squared value, but these declines were not deemed highly influential to change the overall results. Similarly, when recession years were dropped, the coefficient increased marginally. Overall, the robustness check successfully demonstrated that no other factors significantly influenced the results of the benchmark model.

Additionally, heteroskedasticity in the data exists when certain variables, such as STEM major and terminal degree, are unpredictable over time. This unpredictability leads to conditional heteroskedasticity, making it challenging to predict the nature of these variables. Moreover, heterogenous treatment effects can occur when the treated group is subjected to policy change effects within a given time interval, leading to heterogeneity or unpredictability towards certain variables. In this data, the unpredictability of enrolling in a particular STEM major can lead to a heterogeneity effect, as enrollment in the STEM program is the dependent variable, with 1 indicating enrollment and 0 indicating non-enrollment. However, it does not specify the particular major the student is enrolled in, making it unpredictable and displaying unconditional heterogeneity in nature.

By conducting robustness tests and considering potential heteroskedasticity and heterogeneity effects, the study aims to ensure the accuracy and validity of its findings, providing a more comprehensive understanding of the impact of the STEM extension policy on Asian students' enrollment decisions in STEM majors.

The presence of heterogeneity effects is evident in the study, particularly with regards to STEM majors. The results reveal that two specific majors, computer and math sciences, and engineering, are significantly influenced by the STEM extension variable. In other words, the STEM extension policy had a notable impact on the likelihood of enrolling in these majors.

Upon closer examination, it becomes apparent that the engineering major experienced the most substantial impact from the STEM extension policy, followed closely by the computer and math science major. These two fields exhibited the strongest responses to the policy change, indicating that the estimates of the policy's impacts are primarily driven by these two STEM majors.

As a conclusion, the findings suggest that the STEM extension policy had a significant effect on Asian students' decisions to enroll in engineering and computer and math science majors. These two fields demonstrated the highest sensitivity to the policy change, showcasing the importance of the STEM extension in shaping enrollment decisions within these disciplines. This insight provides valuable information for policymakers and educators seeking to promote STEM education and attract international students to specific high-demand fields within the STEM domain.

The STEM extension had a significant impact on Asian students' likelihood of obtaining a STEM MS degree, increasing it by 36% for those who arrived on a student visa. In the case of a Ph.D., the increase was 12%, which was lower than that of the MS degree. From these findings, it can be concluded that the impacts of the STEM extension are primarily driven by the MS and Ph.D. degrees. These results highlight the importance of the STEM extension policy in influencing Asian students' decisions to pursue advanced degrees in STEM fields, particularly at the MS and Ph.D. levels.

**Table 5. Heterogeneity impact of STEM major**

| | Dependent variable: STEM major field | | | | | | |
|---|---|---|---|---|---|---|---|
| No | Variable | Computer and Math Sciences | Life Sciences | Physical Sciences | Social Sciences | Engineering | Science and Engineering Related Fields |
| 1 | STEM extension | 0.038*** | 0.028 | -0.021 | 0.006 | 0.068*** | 0.018 |
| 2 | Age | -0.004 | -0.018 | -0.004 | -0.006 | -0.016 | -0.004 |
| 3 | Age squared | 0 | 0 | 0 | 0 | 0 | 0 |
| 4 | Male | 0.082 | 0.068 | 0.042 | 0.002 | 0.088 | 0.003 |
| 5 | Married | 0.016*** | 0.002 | 0.006 | 0.008 | 0.018** | 0.006 |
| 6 | MS degree | 0.008 | 0.016 | 0.012 | 0.006 | 0.018 | 0.002 |
| 7 | PhD degree | 0.078*** | 0.018 | 0.016 | 0.008 | 0.021*** | 0.003 |
| 8 | R squared | 0.72 | 0.66 | 0.58 | 0.62 | 0.77 | 0.68 |

**Table 6. Heterogeneity impact of terminal degree**

| | Dependent variable: Highest educational degree | | |
|---|---|---|---|
| No | Variable | MS | PhD |
| 1 | STEM extension | 0.366*** | 0.126*** |
| 2 | Age | -0.026*** | -0.032*** |
| 3 | Age squared | 0 | 0 |
| 4 | Male | 0.188*** | 0.262*** |
| 5 | Married | 0.032*** | 0.038*** |
| 6 | Observations | 328,626 | 188,292 |
| 7 | R squared | 0.48 | 0.36 |

## 7. DISCUSSION

The results from the findings support all three hypotheses. Hypothesis 1 is supported by the positive significant impact of STEM extension policy on the Asian student's STEM enrollment decision as shown in equation 1. Hypothesis 2 is supported, where the male is found to be more inclined towards enrolling in STEM programs compared to females. Hypothesis 3 is supported where the option of a doctoral degree is positively affecting the STEM enrollment decision.

The findings from this study support the proposed research design framework and the literature review conducted to develop the framework. In previous literature the study was conducted to understand the effects of STEM policy change in the STEM vs non-STEM major groups, but in this study, we have conducted further analysis by studying effect on STEM policy on immigrant STEM vs US citizen STEM Asian students to explore the possibility of effect of immigration policies changes in STEM provision to tech savvy Asian community, who perceive USA as not only the preferred destination for higher education, but also a destination to permanently migrate. New policies help these Asian students by providing them a pathway to USA citizenship.

**Table 7. Hypothesis results**

| Hypothesis | Result |
|---|---|
| Hypothesis 1 | Supported |
| Hypothesis 2 | Supported |
| Hypothesis 3 | Supported |

## 8. CONCLUSION AND FUTURE DIRECTION

The data from the National Survey of College Graduates, helped this study to examine the effects of STEM extension on STEM major enrollment decisions. This study examined how policy change can affect the enrollment decisions of the Asian ethnic student who arrive in the US on a student visa. The study showed that the STEM extension policy significantly changes the enrollment decision as Asian students leverage the extra 17 months that they get to work in the USA. This also provides an opportunity for them to find a job and get their work permit visa sponsored.

This study only examined a certain phenomenon surrounding Asian student enrollment decisions. Though the enrollment decision is significantly affected by STEM policy the other reason for enrolling in a STEM program are the job opportunities available to the students through the US job market. In future studies, the preference towards IT job roles can be examined from a point of view of STEM policy and student major.

## CONSENT

As per international standard or university standard, Participants' written consent has been collected and preserved by the author(s).

## ETHICAL APPROVAL

As per international standard or university standard guideline participant consent and ethical approval has been collected and preserved by the authors.

## COMPETING INTERESTS

Authors have declared that no competing interests exist.

*Peer-review history:*
*The peer review history for this paper can be accessed here:*
*https://www.sdiarticle5.com/review-history/104323*